\documentclass[conference]{IEEEtran}
\IEEEoverridecommandlockouts
\PassOptionsToPackage{bookmarks=false}{hyperref}
\usepackage{cite}
\usepackage{amsmath,amssymb,amsfonts}
\usepackage{algorithmic}
\usepackage{graphicx}
\usepackage{textcomp}
\usepackage{xcolor}
\usepackage[colorlinks=true,linkcolor=blue,bookmarks=false]{hyperref}%
\usepackage{algorithm}
\usepackage{algorithmic}

\usepackage{booktabs}

\def\BibTeX{{\rm B\kern-.05em{\sc i\kern-.025em b}\kern-.08em
    T\kern-.1667em\lower.7ex\hbox{E}\kern-.125emX}}
\begin{document}

\title{DSSLP: A Distributed Framework for Semi-supervised Link Prediction\\
}

\author{\IEEEauthorblockN{1\textsuperscript{st} Dalong Zhang}
\IEEEauthorblockA{\textit{Ant Financial Services Group} \\
Hangzhou, China \\
dalong.zdl@antfin.com}
\and
\IEEEauthorblockN{2\textsuperscript{nd} Xianzheng Song}
\IEEEauthorblockA{\textit{Ant Financial Services Group} \\
Beijing, China \\
xianzheng.sxz@antfin.com}
\and
\IEEEauthorblockN{3\textsuperscript{rd} Ziqi Liu}
\IEEEauthorblockA{\textit{Ant Financial Services Group} \\
Hangzhou, China \\
ziqiliu@alibaba-inc.com}
\and
\IEEEauthorblockN{4\textsuperscript{th} Zhiqiang Zhang}
\IEEEauthorblockA{\textit{Ant Financial Services Group} \\
Hangzhou, China \\
lingyao.zzq@antfin.com}
\and
\IEEEauthorblockN{5\textsuperscript{th} Xin Huang}
\IEEEauthorblockA{\textit{Ant Financial Services Group} \\
Beijing, China \\
huangxi.hx@antfin.com}
\and
\IEEEauthorblockN{6\textsuperscript{th} Lin Wang}
\IEEEauthorblockA{\textit{Ant Financial Services Group} \\
Beijing, China \\
fred.wl@antfin.com}
\and
\IEEEauthorblockN{7\textsuperscript{th} Jun Zhou}
\IEEEauthorblockA{\textit{Ant Financial Services Group} \\
Beijing, China \\
jun.zhoujun@antfin.com}
}
\IEEEoverridecommandlockouts
\IEEEpubid{\makebox[\columnwidth]{978-1-7281-0858-2/19/\$31.00~\copyright2019 IEEE \hfill} \hspace{\columnsep}\makebox[\columnwidth]{ }}

\maketitle
\IEEEpubidadjcol
\begin{abstract}

Link prediction is widely used in a variety of industrial applications, such as merchant recommendation, fraudulent transaction detection, and so on. 
However, it's a great challenge to train and deploy a link prediction model on industrial-scale graphs with billions of nodes and edges. 
In this work, we present a scalable and distributed framework for semi-supervised link prediction problem (named DSSLP), which is able to handle industrial-scale graphs. 
Instead of training model on the whole graph, DSSLP is proposed to train on the \emph{$k$-hops neighborhood} of nodes in a mini-batch setting, which helps reduce the scale of the input graph and distribute the training procedure. 
In order to generate negative examples effectively,  DSSLP contains a distributed batched runtime sampling module. 
It implements uniform and dynamic sampling approaches, and is able to adaptively construct positive and negative examples to guide the training process. 
Moreover, DSSLP proposes a model-split strategy to accelerate the speed of inference process of the link prediction task. 
Experimental results demonstrate that the effectiveness and efficiency of DSSLP in serval public datasets as well as real-world datasets of industrial-scale graphs.

\end{abstract}


\section{Introduction}
In recent years, graph neural network (GNN), which apply deep neural network to learning representation in graph data, drew lots of attention in the machine learning and data mining community. 
Plenty of literatures demonstrate the effectiveness of applying GNN in solving \emph{link prediction problem} \cite{b44}, which means to predict whether two nodes in a graph are likely to have a link. 
As graph data exists ubiquitously, the link prediction problem has many applications, such as recommendation, fraud detection, knowledge graph completion, etc. 

However, most of GNN-based link prediction models are performed within graphs with millions of nodes and tens of millions of edges, in a single machine. 
There is few literature which focuses on applying GNN-based link prediction models to industral-scale graphs with hundreds of millions of nodes as well as billions of edges, which exist in lots of scenarios among industrial community. 
Take Ant Financial \footnote{https://www.antfin.com/} as an example, there are hundreds of millions customers who pay with Alipay \footnote{https://www.alipay.com/} everyday. 
Billions of transactions are made each month, which form a very-large-scale transaction graph, containing hundreds of millions of nodes (i.e., customers and merchants) and billions of edges (i.e., transactions).  

Many applications, such as personalized merchant recommendation, fraudulent transaction detection, etc., can be formulated as the link prediction problems. 
Take personalized merchant recommendation as an example, transaction between users and merchants can naturally form a user-merchant bipartite graph (i.e., users/merchants as nodes and transactions as edges). 
Hence, personalized merchant recommendation can be formulated as a link prediction problem, which aims to predict whether a certain user would have transactions with a certain merchant in the future, based on the user-merchant transaction graph.

In this paper, we focus on how to build a scalable and distributed GNN-based link prediction model which can be applied to the industrial-scale graphs. This topic is challenged from three aspects. 

First, \emph{scalability}. The proposal need to handle industrial-scale graph data, which may contain hundreds of millions of nodes and billions of edges, as well as the abundant attribute information of nodes and edges. 
In the other hand, to our best knowledge, there is few literature which mentions about link prediction on industrial settings.

Second, \emph{distributed runtime sampling}. To handle industrial-scale graph data, the proposal should perform distributed model training. 
As most conventional GNN-based link-prediction models apply various sampling techniques to generate negative samples \cite{b2,b21}, our proposal should implement effective runtime sampling in the distributed environment. 

Last, \emph{effectiveness of inference}. Different from node prediction (i.e., classification or regression of nodes), the inference procedure of link prediction usually need to perform in much more larger datasets, since the upper bound of the number of links in a graph with $N$ nodes is $O(N^2)$, while $O(N)$ for nodes. 
Therefore, the effectiveness of inference becomes a challenge when facing an industrial-scale graph. 

In this paper, we present a GNN-based \textbf{D}istributed framework for \textbf{S}emi-\textbf{S}upervised \textbf{L}ink \textbf{P}rediction problem, named \textbf{DSSLP}, which is able to perform model training and inference effectively in an industrial-scale graph with billions of nodes and edges. 
First, by avoiding training in the whole graph, we proposed the $k$-hops neighborhood training setting. 
As the representation of a certain node is learned by aggregating the information of its $k$-hops neighborhood, our proposal perform model training in a distributed, mini-batch setting, while each batch contains the subgraph formed by $k$-hops neighborhoods of all nodes in the batch. 
Second, we implement \emph{batched runtime sampling}, which can perform uniform sampling and dynamic sampling, to generate negative samples.
The proposed framework also contains a high-performance inference modules. It reduce the upper bound of computation complexity of node representation inference from $O(N^2)$ to $O(V)$ by model splitting technique. 

The rest of this paper is organizing as follow: 

\section{Related Works}


Link prediction is a long-standing problem that attracts attention from both physical and computer science communities \cite{b8,b9,b10,b11,b12,b13,b48}. 
Various of techniques, such as statistic models, matrix factorization, graph neural networks, and so on, have been applied to this problem. In this section, we will address the most related works that motivate us a lot.


One kind of popular approaches is to evaluate node similarity based on structural information of the graph. 
The basic idea behind is that the more commonality two nodes share in terms of topology, the more likely a link will exist between them. 
As a result, early \emph{heuristic methods} are designed based on different statistics about network structure like degree \cite{b14,b18}, common neighbors \cite{b15,b16,b17}, connected path \cite{b19}. 
Although works well in practice, the performance of these heuristic methods may vary a lot as they usually have strong assumptions on when links exist \cite{b2,b25}. 
Furthermore, researchers also develop algorithms, called \emph{network embedding}, to encode each node of a graph into a low-dimensional vector, and use a certain distance to measure the existence probability of links between node pairs.
A variety of network embedding techniques, such as matrix factorization \cite{b20,b27} , random walk based methods \cite{b11,b12,b22} are engaged into this filed and lead to a superior performance in a number of applications \cite{b23,b24}. 
However,  evaluating node similarity only based on structural information may be limited, especially for nodes only sparsely connected \cite{b21}, or for graphs with abundant attribute information. 
Therefore, they meet their bottleneck when applied to real-world problems \cite{b21}. 


It has been a trend that the link prediction problem refers to not only the structural information, but also node/edge attributes, as the attribute information may also play an important role in deciding whether a node will interact with others in some real-world graphs \cite{b26}. 
\cite{b21} proposes a model linearly combining structural information and attribute information, while \cite{b2,b45} directly encodes structural and attribute information into a low-dimensional feature vector based on \emph{graph neural networks} \cite{b3,b4,b5,b6}. 
\cite{b28} and \cite{b29} state that combining the structural information and attribute information in a graph helps improve performance. 
The main drawback of those methods is scalability, which prevents them from being widely applied to real-world problems.

With the widely use in the industrial applications, these link prediction methods are challenged for their scalability and efficiency. 
Several works aims to alleviate this problem for network embedding methods. \cite{b47} incorporate several modifications to traditional multi-relation embedding systems and present PBG, which is able to handle graphs containing over 100 million nodes and 2 billion edges. 
However, few of work is proposed to leverage both structural and attribute information for link prediction atop the industrial-scale graphs with billions of nodes and edges.

\section{Preliminaries} 
 \label{sec:link_pred_model}
In this section, we will define some notations which will be used in the following sections, as well as describe some algorithmic details of a link prediction model.

\subsection{Notations}
 \label{sec:preliminaries}

%
 
 Generally, a graph can be defined as $G=(V,E)$ with associated adjacency matrix $\mathbf{A}$, where $V$ is the set of nodes $v_i$ and $E$ is the set of edges $(v_i, v_j)$, while $\mathbf{A}_{i,j}=1$ means that there is an edge between node $v_i$ and node $v_j$.
Moreover, many real-world graphs contain node attribute information, which results in the concept of attributed graph (or network)\cite{b46}. 
A real-valued matrix $\mathbf{X} \in \mathbb{R}^{|V| \times m}$ is defined to represent $m$-dimensional node attributes\cite{b1}. 
In attributed graphs, a subgraph not only contains the structural information, but also contains the attributed information.

Here we will introduce a concept named \emph{$k$-hops neighborhood}. 
The $k$-hops neighborhood of node $v_i$ (denoted as $\mathcal{N}^{(k)}_{v_i}$) in an attributed graph $G$ is defined as a subgraph induced from $G$ by a set of nodes $\{ v_j | d(v_i, v_j) \leq k, v_j \in V\}$, where $d(v_i, v_j)$ denotes the length of shortest path from node $v_i$ to node $v_j$.
 
 \subsection{Node Embedding} 
 \label{sec:per_node_embedding}
Node embedding (or called graph embedding) methods aim to encode nodes into low-dimensional vectors, which is able to preserve some kinds of graph properties. 
One kind of popular methods to learn a deep encoder is to update parameters by aggregating information from the local neighborhood of a node in an iterative fashion \cite{b3,b4,b5,b6}. Hence, the embedding of node $v_i$ in $(k+1)$-th hidden layer is denoted as:
 \begin{equation}
 \label{eq:df_node_embedding}
\mathbf{h}^{(k+1)}_{i} = \phi(\{\mathbf{h}^{(k)}_{j} | v_j \in \mathcal{N}^{(1)}_{v_i}\}; \mathbf{W}^{(k)})
 \end{equation}
 where $\phi(\cdot)$ is an aggregator function (i.e., mean, sum and max pooling operator, and so on) defined on the set of $k$-th hidden layer embeddings of its 1-hop neighborhood, $\mathbf{W}^{(k)}$ denotes the trainable parameters of $k$-th hidden layer, and $\mathbf{h}^{(0)}_{i} = \mathbf{X}_i$. 
By stacking the aggregating operator $K$ times, each node embedding is able to aggregate information from its $K$-hops neighborhood.

\subsection{GNN-based Semi-supervised Link Prediction Model}
 \label{sec:model}
Given a snapshot of a graph, the link prediction problem aims to accurately predict the links (edges) that will be added to the graph in the future. 
It naturally follows a semi-supervised setting, in which existing links in the graph are treated as positive examples, while the negative examples can be generated by different sampling strategies.

Formally, from an encoder-decoder perspective \cite{b1}, the existence score of a link between node $v_i$ and node $v_j$ is defined as:
\begin{equation}
\label{eq:linkpred_en_dec}
\mathcal{S}(v_i, v_j) = f_{dec}(f_{enc}(v_i), f_{enc}(v_j))
\end{equation}
where $f_{enc}:V \to \mathbb{R}^d$ is a function that maps node set $V$ to $d$-dimensional embedding space $\mathbb{R}^d$, while $f_{dec}:\mathbb{R}^d \times \mathbb{R}^d \to \mathbb{R}$ is a pairwise decoder that maps pairs of node embeddings to a real-valued edge score measurement. 

Usually, the encoder $f_{enc}$ and decoder $f_{dec}$ can be learned by optimizing an emprical reconstruction loss:
\begin{equation}
\label{eq:emprical_loss}
\mathcal{L}= \sum_{(v_i, v_j) \in \mathcal{D}}{\ell(\mathcal{S}(v_i, v_j), \mathcal{G}(v_i, v_j))}
\end{equation}
where $\mathcal{G}(v_i,v_j)$ is a user-specified pairwise proximity measure. Specially, $\mathcal{G}(v_i,v_j)=1$ represents that there is a link between node $v_i$ and $v_j$, while $\mathcal{G}(v_i,v_j)=0$ indicates no link exists. 
Meanwhile, $\ell(\cdot)$ represents a user-specified loss function (e.g. $L_2$ distance) measuring the discrepancy between estimated value $\mathcal{S}(v_i, v_j)$ and the ground truth $\mathcal{G}(v_i, v_j)$, and $\mathcal{D}$ is a dataset for model training.

In the GNN-based link prediction model,  a deep encoder with $K$ layers (i.e., Equation \ref{eq:df_node_embedding}) is applied to generate node embeddings. 
The decoder $f_{dec}(\cdot)$ here can be dot product or euclidean distance of two node embeddings, or a classifier (i.e., logistic regression and multilayer perceptron) defined on the concatenation of two node embeddings. That is 
\begin{equation}
\label{eq:inner_prodcut}
\mathcal{S}(v_i, v_j) = f_{dec}(\mathbf{h}^{(K)}_{i}, \mathbf{h}^{(K)}_{j})
\end{equation}

$\mathcal{S}(v_i, v_j)$ is used to measure the existence score of an edge between nodes $v_i$ and $v_j$. The larger the score $\mathcal{S}(v_i, v_j)$ is, the more likely that an edge will appear between node $v_i$ and $v_j$. To that end, we penalize a margin-based ranking loss \cite{b31,b32} to guide the learning process:
\begin{equation}
\label{eq:ranking_loss}
\mathcal{L} =\sum_{(v_i,v_j) \in E,(\acute{v_i}, \acute{v_j}) \sim \mathcal{P}}max(\mathcal{S}(\acute{v_i}, \acute{v_j}) - \mathcal{S}(v_i, v_j) + \lambda, 0)
\end{equation}
where $(v_i,v_j) \in E$ represents a positive edge while $(\acute{v_i}, \acute{v_j})$ is a negative example sampled from a certain distribution $\mathcal{P}$. 
In our model, positive and negative edges are sampled according to a set of sampling strategies, which will be detailed in the following section.
 $\lambda$ is a margin hyperparameter to separate positive and negative edges. 
The loss function in \eqref{eq:ranking_loss}  ensures that negative edges get lower scores than positive edges by at least a margin $\lambda$. 
In addition, we also provide a classification loss as an alternative. 

%

By minimizing the loss function introduced above via mini-batch stochastic gradient descent (SGD), DSSLP jointly learning node embedding $\mathbf{h}_i$ and link prediction score $\mathcal{S}(v_i, v_j) $. Therefore, DSSLP can be not only used in link prediction task, but also transferred to other tasks like node classification task with few extra labeled data.

\section{Framework}
 \label{sec:framework}
\subsection{Overview}
\begin{figure}[htbp]
  \centering
  \includegraphics[width=0.95 \linewidth]{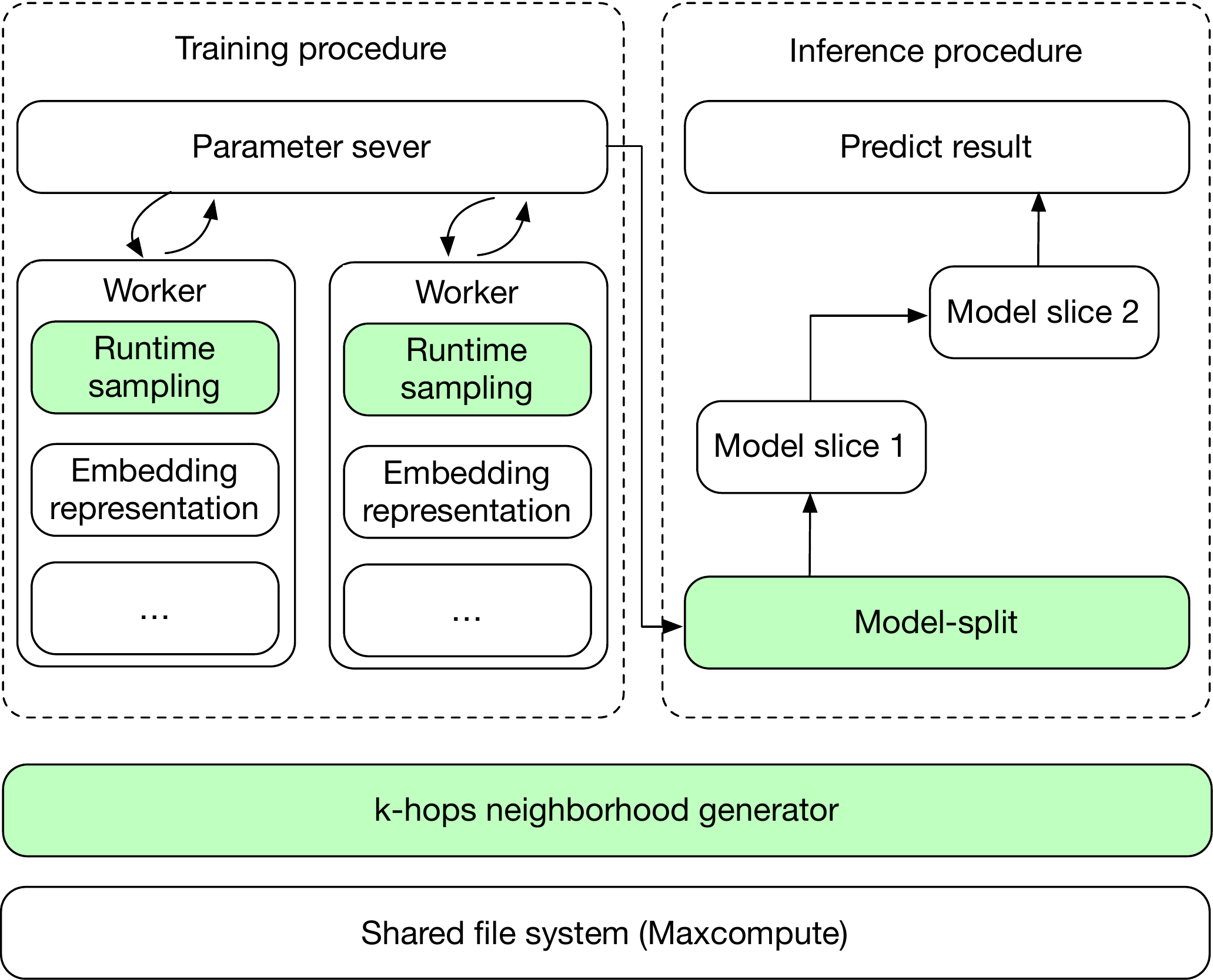}
  \caption{Overview of DSSLP Framework}
  \label{fig:dist_frame}
\end{figure}

DSSLP is designed to solve industrial-scale link prediction problems. As illustrated in Fig.~\ref{fig:dist_frame}, DSSLP provides a complete solution for link prediction applications: (1) Data preprocessing. Generate $k$-hops neighborhood for each node and store them into a shared file system (Maxcompute \footnote{https://www.alibabacloud.com/product/maxcompute}); (2) Training. Train the link prediction model described above in a distributed and asynchronous manner based on the parameter-sever architecture \cite{b30}; (3) Inference. Compute link prediction score for industrial-scale graph data with carefully designed model-split strategy. 

\begin{figure*}[htbp]
  \centering
  \includegraphics[width=0.95 \linewidth]{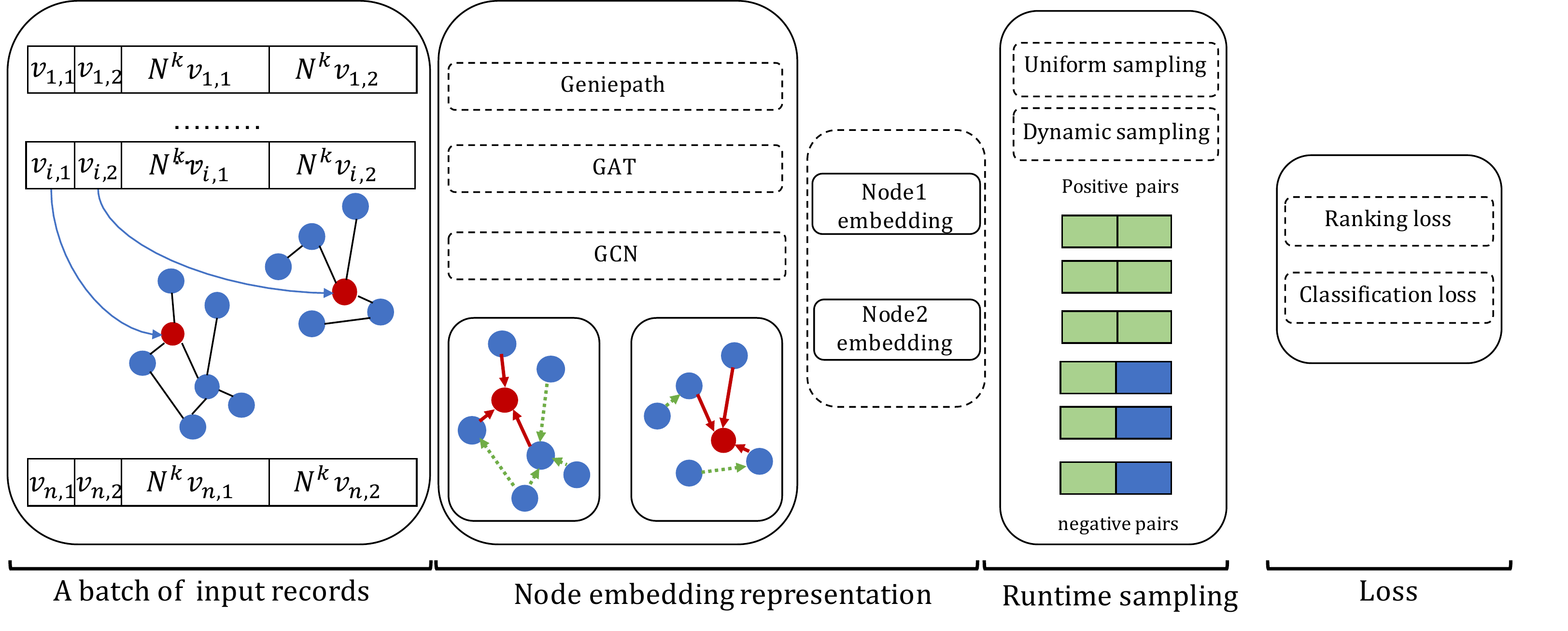}
  \caption{Work flow of DSSLP framework durning training time.}
  \label{fig:work_flow}
\end{figure*}

DSSLP receive a special format of \emph{$k$-hops neighborhood} as input rather than the full graph, which helps reduce the time and memory cost. During training procedure, every worker pulls parameters from a parameter sever and update them independently. For each worker, we present the work flow in Fig.~\ref{fig:work_flow}, which mainly consists of three parts: (1) Encode node embeddings based on subgraphs with a certain of embedding representation algorithms like GCN \cite{b3}, GAT \cite{b5}, Geniepath \cite{b6}; (2) Sample positive and negative links according to observed links within a batch; (3) Optimize a certain loss to guide the learning process. After that, we can apply a well-trained model to predicting the existence of links in a graph. Specially, for industrial-scale graph data, we split the well-trained model into two parts to avoid some repeated computing and accelerate the inference procedure. 



When applied to industrial-scale link prediction applications, a framework should not only care about performance, but also hold the properties of high quality in scalability and efficiency. In the following, we will detail our framework and explain how this framework is possible to efficiently and accurately predict links between arbitrary two nodes on huge graphs.


\subsection{Operating on $k$-hops neighborhood}
The first challenge for a link prediction model is how to operate on huge graphs with billions of nodes and edges. The cost of memory and time will arrive at an unacceptable level when loading a full huge graph into memory, as a typical graph that contains one billion nodes with each associated with 64-dimensional float features would require 250G memory to just store these features,  which is beyond the capacity of typical commodity machines. 


Rethinking the total work flow of the link prediction model, we find that, although embedding representation is based on structure and attribute information of a graph, it's unnecessary to load the full graph into memory at once. As stated in section ~\ref{sec:per_node_embedding}, embedding a node is to aggregate information from its local neighborhood and a $k$-hops embedding of node $x$ can be accurately calculated from its $k$-hops neighborhood $\mathcal{N}^{(k)}_{v_i}$ \cite{b2}. Therefore, we format our inputs based on $k$-hops neighborhood instead of the full huge graph.

Our DSSLP framework takes pairs of nodes together with their $k$-hops neighbor as input. As shown in Fig.~\ref{fig:work_flow}, the input of our framework is encoded into an ``edge'' format, which mainly consists of four components: (1) Id of node $v_1$; (2) Id of $v_2$; (3) $N^{(k)}_{v_1}$, $k$-hops neighborhood of node $v_1$; (4) $N^{(k)}_{v_2}$, $k$-hops neighborhood of node $v_2$.  Ids of $v_1$ and $v_2$ are used to label the root nodes of subgraphs and the node pair $(v_1, v_2)$ forms a link to be predicted, while $N^{(k)}_{v_i}$ contains information of nodes and edges of the $k$-hops neighborhood rooted in node $v_i$. 

For a certain node $v_i$, the $k$-hops neighborhood $N^{(k)}_{v_i}$ mainly consists of two parts: nodes associated with node features and edges together with their features in $k$-hops neighborhood of $v_i$, which is constructed by the following steps:
\begin{enumerate}
\item Insert the root node $v_i$ into the node list $\mathbf{T_n}$ together with its features.
\item For each node in $\mathbf{T_n}$, insert edges directly connected to them into the edge list $\mathbf{T_e}$
\item Update $\mathbf{T_n}$ with nodes associated with edges in $\mathbf{T_e}$.
\item Repeat the second and third steps for $k$ times.
\end{enumerate}
By alternately update the node list $\mathbf{T_n}$ and the edge list $\mathbf{T_e}$, we finally generate the $k$-hops neighborhood for node $v_i$.

To prepare the input of our model, we first extract $k$-hops neighborhood of $v_1$ and $v_2$ respectively, and then construct input with the ``edge'' format  stated above. 
Since the $k$-hops neighborhood is information complete for computing the $k$-hops embedding of its root node, this kind of input provide sufficient structural and attribute information for our model to predict the existence between nodes $v_1$ and $v_2$.


With input in such format, our DSSLP model is born with the ability of running parallelly and distributedly, as the model is able to learn embeddings of root nodes directly from a piece or pieces of data records without the need of referring to the full graph. 
Furthermore, our model is memory friendly, as the scale of input is decreasing from a full graph to a $k$-hops subgraph. 
For example, given a graph with 0.6 billion nodes and 10.4 billion edges, the average node degree is about 17, and then a $3$-hop subgraph $N^{(3)}_{v_i}$ only contains about $17^3=4913$ nodes, which is far less than the amount of nodes in the full graph.

\subsection{Batched Runtime Sampling}

The second challenge is how to efficiently sampling negative edges to guide the learning process. 
A straightforward way is sampling negative links in an offline manner: randomly select a certain number of observed edges as positive links together with about the same number of unobserved edges as negative links \cite{b2,b21}. 
However, when applied into industrial-scale applications, offline sampling may only cover a small part of the whole possible negative links and leads to an ``imbalance" problem \cite{b21}.
What's worse, the imbalance data may break the basic assumption that training and test set should obey the same distribution. 
To solve this problem, we propose a batched runtime sampling method, which is flexible for different scale of graphs and is able to approximate the distribution of positive and negative edges of the whole observed graph $G$.


The runtime sampling method operates on a batch of records during the training procedure. 
Given a batch of records, a pair of root nodes $(v_{i,1},v_{i,2})$ in the $i$-th record is sampled from the whole observed edge set $E$, which naturally forms a \textbf{positive link}. 
Meanwhile, a pair of nodes $(v_{i,j},v_{m,n})$ is sampled as a \textbf{negative link} if $v_{i,j}$ and $v_{m,n}$ are root nodes within the batch and the edge $(v_{i,j},v_{m,n})$ can not be observed in the edge set $E_b$ (all edges within the batch).
As all nodes directly connected to the root node $v_{i,j}$ are collected in $v_{i,j}$'s $k$-hops neighborhood, sampling negative links according to $E_b$ is equal to sampling according to the whole edge set $E$. 

A set of strategies are developed to make our runtime sampling method flexible and powerful: 

\paragraph{Shuffling}  
One key point for batched runtime sampling is that negative examples should be sampled according to the whole dataset rather than only within a batch.
While candidates of negative links are combinations of nodes from the whole graph, batched sampling makes candidates of negative samples be limited to the combinations of root nodes within a training batch, which may not reflect the real distribution of the whole graph. 

A simple but efficient shuffling strategy is used to solve this problem when we design our batched runtime sampling method. 
The basic idea behind is that we should provide enough combinations of nodes as candidates of negative samples to sampling methods, which, is equal to provide enough combinations of data records. 
To that end, we perform a runtime shuffling strategy: First, we pick $m$ records \emph{in order} from the training dataset and put them into a prefetch buffer;
Then, we \emph{randomly} pick $n$ records one by one to fill a training batch. 
Note that, once a record was pulled from the buffer, a new record will be picked from train dataset and replace the former record in the buffer. 

With such shuffling strategy, a training batch is filled with randomly selected records, and a certain data record will meet any other records with enough training steps. Therefore, the candidates of negative links generated by batched runtime sampling approximate those sampled over the whole training dataset.

\paragraph{Uniform sampling}  We provide a basic uniform negative sampling method in our sampling suite. 
The uniform negative sampling method is designed based on the basic assumption that every node plays an equal role in the graph. 
Therefore, the uniform negative sampling method treats every node equally without any bias, and samples about the same number of negative neighbors from each node to form negative links. 

The uniform negative sampling method takes a batch of records as input and output a set of negative links according to root nodes and observed edges within the training batch. Given a root node $x$ from a certain record, the uniform sampling method works as follows:
\begin{enumerate}
\item Randomly select a root node $y$ from the training batch.
\item Pick the link $(x,y)$ as a negative link if it can not be observed in the edge set $E_b$ of the batch, and drop otherwise.
\item Repeat first two steps until getting $neg\_num$ negative links or reaching pre-defined $max\_trail$ times.
\end{enumerate}

It's worth noting that, even without the shuffling strategy, the uniform negative sampling method is also able to provide a diversity of negative examples for a certain node, as negative links are randomly constructed. 
The uniform negative sampling method is just a basic solution for runtime sampling and it can not approximate the real distribution in some real-world problems, as the importance of a node may vary due to different roles it played in the real-world applications.
It is a good choice to apply this basic sampling method to applications where we don't have any prior knowledge, and even statistic informations like in-degree, out-degree, are not available or beyond belief.

\paragraph{Dynamic sampling} Moreover, we also propose a dynamic sampling method based on the degree of each node to approximate the distribution of the whole observed graph $G$. 
It is a common sense that the larger the degree of a node is, the more likely we will observe a link connected to it.
Therefore, we form an inverse relationship between node degree and the number of negative links associated with the node $v_i$:
\begin{equation}
\label{eq:inverse_relation}
n_{v_i} =  \frac{\eta}{\lambda_{v_i}}
\end{equation}
where $n_{v_i}$ is the number of negative edges to be sampled,  $\lambda_{v_i}$ represent the degree of node $v_i$, and $\eta$ is a ratio hyperparameter to adjust the value of $n_{v_i}$. In case that $\lambda_{v_i}$ is extremely large or small, we add two extra parameters $t$, $\alpha$ to \eqref{eq:inverse_relation}:
\begin{equation}
\label{eq:inverse_relation_additional}
n_{v_i} = \left \{
\begin{array}{ll}
max (\lfloor\frac{\eta}{\lambda_{v_i}}\rfloor, 1), & \lambda_{v_i} > t \\
min(\lfloor \frac{\eta}{\lambda_{v_i}}\rfloor, \alpha), & \text{otherwise}
\end{array}
\right.
\end{equation}
where $\alpha$ is the max negative number to be sampled for a node. $t$ is a threshold to adjust the real sampling number. \eqref{eq:inverse_relation_additional} ensures that the number of negative links for a certain node should be in the range of $[1, \alpha]$.

Different from the uniform sampling method described above, our dynamic sampling method adjusts the number of negative edges associated with a certain node dynamically according to \eqref{eq:inverse_relation_additional}. It worth noting that, from a statistic perspective, node degree is equal to the frequency of positive links connected to the node, which naturally reflects the real distribution of observed links. Therefore, examples sampled by our dynamic sampling method will approximately obey the same distribution of links in the observed graph $G$, as the sampling procedure takes the node degree into account.


\subsection{Offline Inference}
The third challenge is that, as the total number of links to be predicted may reach a quite large scale, as many as billions or even trillions,
 it may take tens or hundreds of hours to perform the inference over an industrial-scale dataset, which is usually beyond tolerance for industry applications.
To solve this problem, we propose a model-split strategy to accelerate the inference procedure. 

\begin{figure*}[htbp]
  \centering
  \includegraphics[width=0.90 \linewidth]{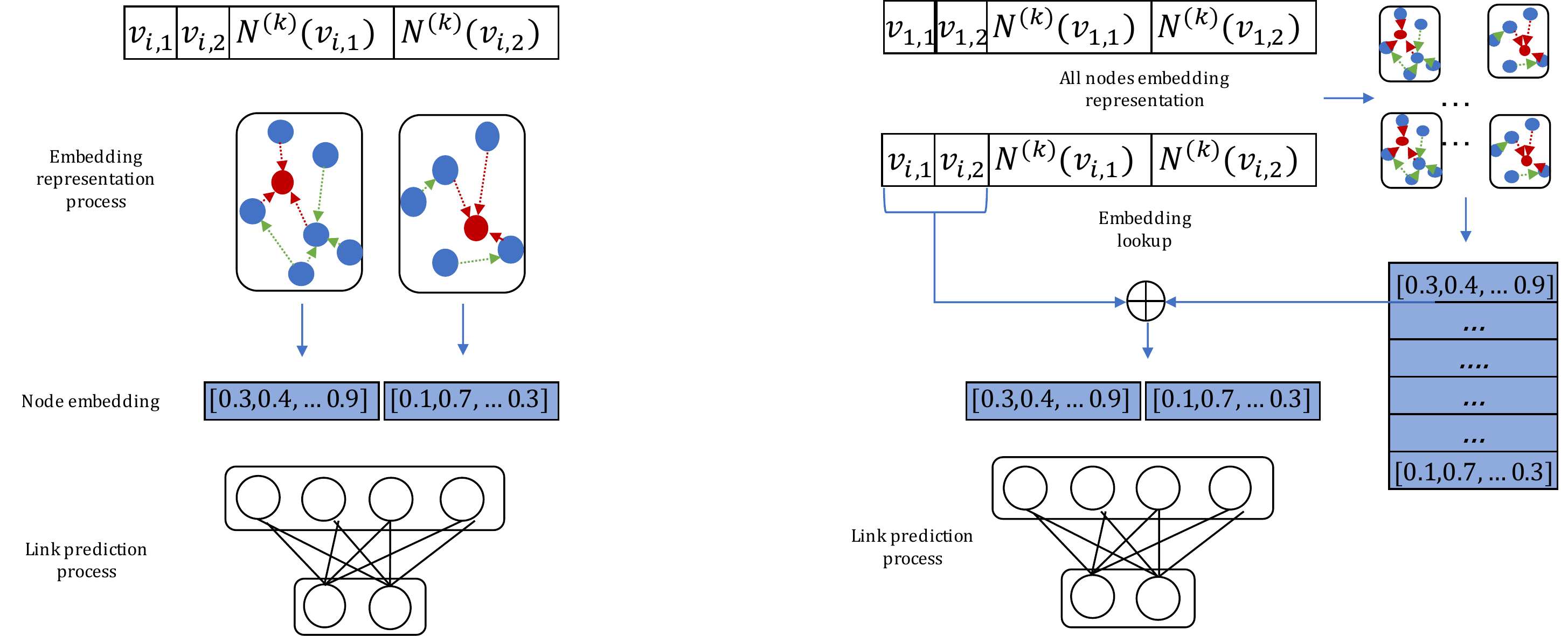}
  \caption{Original inference process VS. Inference with model-split strategy.}
  \label{fig:model_split}
\end{figure*}

The basic idea of the model-split strategy is to reduce the computational complexity of the inference procedure.
Given a graph $G$ with $N$ nodes, the upper bound of the number of links is $O(N^2)$, while $O(N)$ for nodes, which means that we should perform the forward propagation procedure of our model as many as $N^2$ times. 
However, there are many unnecessary computations, as the same node may appear in different edges and its embedding is computed repeatedly in different input records, which leads to a high time cost.
Based on this observation,  we split the inference procedure into two parts: node embedding representation procedure and prediction procedure. As shown in Fig.~\ref{fig:model_split}, The model-split strategy mainly consists of four steps:
\begin{enumerate}

\item Split parameters of well trained model into two parts: parameters related to embedding representation procedure and prediction-related parameters.
\item Compute node embeddings with parameters related to embedding representation procedure.
\item Pack node embeddings into the edge format with reference to node ids from edge candidates.
\item Compute prediction scores with prediction-related parameters.
\end{enumerate}
Note that if we use ranking loss during training procedure, there's no prediction-related parameters, as we directly use the result of dot product as the prediction score. If we use the alternative classification loss, the prediction score is computed by a neural network and all the four steps should work in the inference procedure.

Compared with the original inference procedure, the model split strategy reduce the upper bound of the computational complexity in embedding representation procedure from $O(N^2)$ to $O(N)$.  As embedding representation is the most time consuming part of our model, the model-split strategy is efficient to speed up the inference procedure. 


\section{Experimets}
In this section, we present the experimental results of our approaches on some public datasets together with some real-world applications at Alipay.

\subsection{Datasets}
\label{sec:dataset}
The datasets we used mainly consist of two parts: public datasets and some real-world industrial-scale data at Alipay.

\paragraph{Public Datasets} Following \cite{b2,b27}, we evaluate our approaches on seven different datasets :

\begin{itemize}
\item USAir \cite{b33}: a network of US air transportation system with 332 nodes and 2126 edges. Each edge represents an airline from source city to destination.   
\item NS \cite{b34}: a co-authorship network of researchers in network science field with 1589 nodes and 2742 edges.
\item PB \cite{b35}: a network of the US political blogs with 1222 nodes and 16714 edges.
\item Yeast \cite{b36}: a network of molecular interactions in cells of Yeast with 2375 nodes and 11693 edges.
\item C.ele \cite{b37}: a neural network of C. elegans with 297 nodes and 2,148 edges.
\item Power \cite{b37}: an electrical power grid of the western US with 4941 nodes and 6594 edges.
\item Router \cite{b38}: a router-level Internet collected by Rocketfuel Project with 5022 nodes and 6253 edges.
\end{itemize}

Details about those dataset are presented in Table \ref{tab:summary_open_data}. $node\_degree$ means the average node degree of a dataset and implies the sparsity of related network. 

\begin{table}[htbp]
\caption{Summary of public datasets}
\begin{center}
\label{tab:summary_open_data}
\begin{tabular}{c||ccc}
\toprule
Indices &Vertices &Edges & node\_degree \\
\midrule
USAir   &332           &2126      &12.81    \\
NS       &1589         &2742      &3.45   \\
PB       &1222         &16714    &27.36   \\
Yeast   &2375         &11693    &9.85    \\
C.ele   &297           &2148      &14.46    \\
Power &4941         &6594      &2.67  \\
Router&5022         &6258      &2.49   \\
 \bottomrule
\end{tabular}
\end{center}
\end{table}

\paragraph{Real-world Dataset At Alipay}  We apply our approaches to real-world problems at Alipay, the world's largest mobile payment platform. Datasets we used can be concluded as follows:

\begin{itemize}
\item Trade Network. Trade network describes transactions between different users (i.e., customers and merchants). In the trade network, each user is treated as a node and we add an edge to the network if a deal happens between two users. We preprocess the data by deleting isolated nodes which are useless in propagating information through the topology \cite{b6}. The preprocess data contains about 376 million nodes and 4.48 billion edges. Moreover, a set of attributes is used to describe users and edges. Each node is associated with 3001-dimensional features while each edge contains 2-dimensional features. A few nodes in trade network are labeled to identify a special role in transaction. 

\item Friendship Network. Friendship network describes the friendship between different individuals on the Internet. After preprocessing the data as stated above, the friendship network contains about 638 million nodes each with 446-dimensional features and 10.4 billion edges with 28-dimensional features. As time goes on, two individuals may build a friendship and the dataset records the start time of a friendship with a timestamp, which is used to separate the whole dataset into train and test sets.
\end{itemize}

\begin{table}[htbp]
\caption{Summary of Alipay Datasets}
\begin{center}
\label{tab:summary_aipay_data}
\begin{tabular}{c||cc}
\toprule
Indices &Trade Network &Friendship Network  \\
\midrule
\#Nodes   &$3.76\times 10^8$           &$6.38\times 10^8$         \\
\#Edges       &$4.48 \times 10^9$         &$1.04 \times 10^{10}$       \\
\#Node feature       &3001         &446       \\
\#Edge feature   &2         &28        \\
Task type(test) & node classification &link prediction \\
label type(test) &node label &edge label \\
\#Train set(link) & $1.14 \times 10 ^8$              & $1.02 \times 10^8$\\
\#Test set (link) & $-$        &$6.2 \times 10^6$\\
\#Train set (Node) &$6.96 \times 10^6$  &$-$\\
\#Test set (Node)&$1.04 \times 10^6$         &$-$        \\
 \bottomrule
\end{tabular}
\end{center}
\end{table}

To prove the effectiveness of our approaches, which jointly learns link prediction and node embedding, we preform a node classification task on Trade Network and a link prediction task on Friendship Network.

For the node classification task on Trade Network, we first train a link prediction model and then feed node embeddings learned by the model to downstream node classification task with provided node labels. 
We present details about train and test set in Table \ref{tab:summary_aipay_data}. About $1.14 \times 10^8$ training records are used to train the link prediction model while $6.96 \times 10^6$ labeled node are used to train the classifier. It's worth noting that, extra node labels are only used to train the classifier and play no role in node embedding representation procedure. 

On the other hand, for the link prediction task on Friendship Network, we first use a timestamp to split the graph into two part: links built before the timestamp as observed links while links after the timestamp as the positive test set. As for the negative links in test set, we randomly sampled unobserved links according to the graph. The total training set contains about $1.02 \times 10^8$ records, while the test set contains $6.2 \times 10^6$ records with about equal number of positive and negative links. It's worth noting that all the training records are built based on observed links, and we use the runtime negative sampling method to generate negative samples during training time. Therefore, there is no need to prepare negative samples for training set.

\begin{table*}[htbp]
 \caption{Results on public dataset}
 \begin{center}
 \label{tab:result_open_dataset}
 \begin{tabular}{c||ccccccccccccccc}
  \toprule
  Algorithm  & USAir    &{}  & NS     &{}    & PB    &{}   & Yeast   &{}   & C.ele   &{}    & Power   &{}      & Router &{} & Mean\\
  \midrule
  CN          & 0.940   &{}    & 0.938  &{}   & 0.919   &{} & 0.891    &{}  & 0.848   &{}    & 0.590    &{}      & 0.561  &{}  & 0.812\\
  Jaccard  &0.903     &{}   & 0.938  &{}    & 0.873   &{} & 0.890    &{}  & 0.792   &{}    & 0.590    &{}       &0.561  &{}  &0.792\\
  AA          &0.950     &{}  &0.938     &{}   & 0.922   &{} &0.891     &{}   & 0.864   &{}     & 0.590   &{}       &0.561  &{}  & 0.817\\
  RA         &0.956      &{}  & 0.938   &{}   & 0.923    &{} & 0.892    &{}   &0.868    &{}    & 0.590    &{}      &0.561   &{}  & 0.818 \\
  N2V      &0.914       &{}  & 0.915   &{}   & 0.858   &{}  & 0.937    &{}   & 0.841   &{}    & 0.762    &{}      & 0.655  &{}  & 0.840\\
  WLMN  &0.961       &{}  & 0.981   &{}   &\underline{0.939}    &{}  &0.951     &{}   & 0.854  &{}     &0.874     &{}      & 0.915  &{}  & 0.925\\
  SEAL    & \underline{0.966}      &{}  &\underline{0.989}   &{}   & \underline{\textbf{0.947}}  &{}  &\underline{\textbf{0.979}}  &{}     & \underline{\textbf{0.903}}   &{}   & \underline{0.876}    &{}      &\underline{0.964} &{} &\underline{ 0.946}\\
  \midrule
  Ours     &\underline{\textbf{0.982} } &{}     &\underline{\textbf{0.992}} &{}   &0.923  &{}  &\underline{0.974}  &{}   &\underline{0.885}     &{}   &\underline{\textbf{0.921}}      &{}  &\underline{\textbf{0.970}}  &{}&\underline{\textbf{0.950}}\\
  
   \bottomrule
 \end{tabular}
 \end{center}
\end{table*}

\subsection{Experimental Settings}

We evaluate the performance of our model mainly on two tasks: the link prediction task and the node classification task. Experimental Settings are designed as follows:

\begin{itemize}
\item Link prediction task. We first compare our model with some popular methods in academic, including : Common Neighbors (CA) \cite{b15} , Jaccard index (Jaccard) \cite{b39},  Adamic-Adar (AA) \cite{b16},  resource allocation (RA) \cite{b17}, node2vec \cite{b12}, Weisfeiler-Lehman Neural Machine (WLNM) \cite{b27}, SEAL \cite{b2}. The first four methods are heuristic methods while the last three represent latest learn-based methods. We report results of our methods together with those methods on seven public dataset introduced in section \ref{sec:dataset}. Furthermore, we conduct a set of experiments on Friendship dataset with node2vec \cite{b12} as the baseline, which, to our best knowledge, is a public available method that is able handle such large scale graph.

\item Node classification task. As our model jointly learn link prediction and node embedding, we apply the learned embedding to node classification task on Trade dataset. Also, we use one of the most famous unsupervised embedding representation method node2vec\cite{b12} as the baseline. Gradient Boosting Decision Tree \cite{b40} (GBDT) is adopted as the classifier which is widely used in industry.
\end{itemize}

We report the AUC measure on all those experiments, because it is not influenced by the distribution of the classes compared with 0-1 accuracy. Moreover, we use Geniepath \cite{b6} as the default embedding representation method and embedding size of our model is set to 64 in all of those experiments. When performing the link prediction task, inner-product is used as the metric function of our model. 






\subsection{Results}
Following the evaluation protocol stated above, we details results on public datasets and read-world industrial-scale datasets in terms of performance in this section.

\paragraph{Link prediction task} We report results in terms of link prediction task in Table \ref{tab:result_open_dataset} and Table \ref{tab:result_alipay_dataset}. In table \ref{tab:result_open_dataset}, we compare our model with some popular methods in academic on seven public dataset. \underline{Underline} is used to indicate the top-2 results and \textbf{Bold} represents the best results. Meanwhile, we illustrate results on real-world industrial-scale dataset in Table \ref{tab:result_alipay_dataset}.


As shown in Table \ref{tab:result_open_dataset}, our methods consistently outperforms stated heuristic methods like Common Neighbors (CA) \cite{b15}, Jaccard index (Jaccard) \cite{b39}, Adamic-Adar (AA) \cite{b16},  resource allocation (RA)  \cite{b17} along all seven public dataset. On Router dataset, our method even achieve an  AUC of 0.970 while those heuristic methods only get 0.561. And the mean AUC across the seven dataset of our method is 0.950 while the best mean AUC of those four heuristic methods is 0.818. In all, our method achieves a significant improvement in terms of AUC compared with those heuristic methods.

Our methods also show advantages over some latest learning-based methods such as node2vec (N2V) \cite{b12}, WLMN \cite{b27}, and SEAL \cite{b2}. Compared with the state of the art method, SEAL, our model win four out of seven dataset in terms of AUC and the performance on the left three dataset is also about the same level with that of SEAL. Results shown in Table \ref{tab:result_open_dataset} demonstrate that our model achieves competitive performance compared with those latest methods in academic on small public datasets. 

Moreover, results on large-scale dataset also demonstrate the effectiveness of our model.
Note that, when applied to large-scale dataset, prior methods may fail to work due to the high cost of memory. 
As a result, we mainly take node2vec as the baseline, which, to our best knowledge, is a public available method to handle such large graph and also widely used in industrial applications. 
As presented in Table \ref{tab:result_alipay_dataset}, our model outperforms node2vec by about $15\%$ (0.842 VS. 0.738). As node2vec only embed the structure information without referring to attribute information, we also show the result based on node2vec embedding with raw features for fairness. Our model also achieves significant improvement compared with node2vec + raw features(0.842 VS. 0.769).  

\paragraph{Node classification task} Results in terms of node classification task on Trade dataset are presented in Table \ref{tab:result_alipay_dataset}. As our model learns node embeddings from such large scale dataset without referring to any node labels, we also use node2vec as the baseline.

The node embedding of our method is more distinctive compared with that of node2vec as shown in Table \ref{tab:result_alipay_dataset}. As the node labels in trade dataset represent a special role in transactions, which may mainly be influenced by attributes of a node, embeddings with structural information only may not works well. Therefore, node2vec achieves only 0.569 in terms of AUC on Trade dataset, which is far less than that of our model. Furthermore, our model outperforms node2vec significantly (0.757 VS. 0.658),  even node2vec is combined with raw features.



\begin{table}[htbp]
\caption{Result on Real-world Dataset At Alipay}
 \begin{center}
  \label{tab:result_alipay_dataset}
 \begin{tabular}{ccc}
  \toprule
  Name & Trade (Node)          & Friendship (Link) \\
   \midrule
   raw feature & 0.639 & 0.702 \\
   node2vec              & 0.569 & 0.738 \\
   node2vec + raw feature    & 0.658    & 0.769\\
   Ours            & \textbf{0.757}& \textbf{0.842} \\
  \bottomrule
   \end{tabular}
    \end{center}
\end{table}

\subsection{Analysis}

\begin{figure}[htbp]
\begin{minipage}[t]{0.50\textwidth}
\centering
\includegraphics[width=0.99\textwidth]{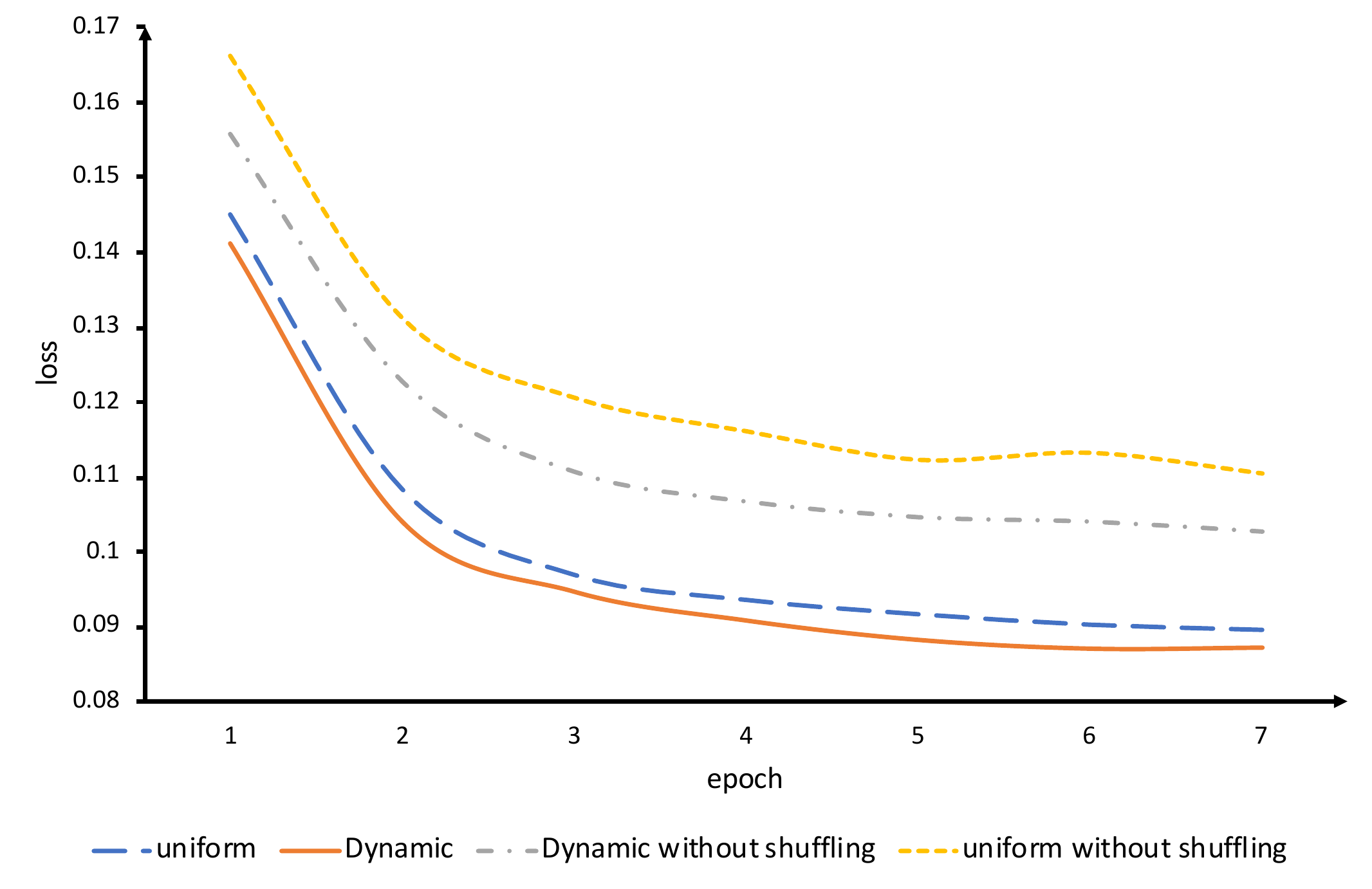}
\caption{Loss curve varying according to different strategies}
\label{fig:loss_uniform_vs_indegree}
\end{minipage}%
\end{figure}

In this section, we try to analysis and explain the effect of different strategies used in our model by varying the variable we will analysis while fix other variables which is described as follows:


\paragraph{effectiveness} We analysis the influence of shuffling strategy and sampling strategy to show the effectiveness of those strategies. Results of all four combinations across those two strategies are illustrated in Fig.~\ref{fig:loss_uniform_vs_indegree}, in which ``Uniform" and ``Dynamic" refer to model with uniform sampling and dynamic sampling respectively, while suffix ``without shuffling" indicates model trained without shuffling strategy. The horizontal axis means the training epoches while vertical axis means losses.

As illustrated in Fig.~\ref{fig:loss_uniform_vs_indegree}, both shuffling strategy and dynamic strategy are proved to be effective to help minimize the loss to a better place. By fixing the sampling strategy, we can conclude the effect of the shuffling strategy: when enabling the shuffling strategy, both loss curves of models with uniform sampling and dynamic sampling decrease to a lower place compared with models without shuffling strategy. On the other hand, by fixing shuffling shuffling strategy, models with dynamic sampling strategy achieve better result than models with uniform sampling. Moreover, model with dynamic negative sampling and shuffling strategies achieve best result across the four models.

The result described above may be due to the reason that, by shuffling, a certain record is able to meet any other records across the whole training set, which is equal to train the model on the whole dataset rather than within a min-batch. Moreover, by adopting dynamic negative sampling, the distribution of positive and negative links in training set approximates that over the whole observed graph, which helps construct more reasonable number of positive and negative samples.


\paragraph{Scalability}  For link prediction tasks on industrial-scale dataset, scalability is an import dimension that we should take into account. To prove the scalability of our framework, we design a set of link prediction experiments by varying worker numbers during training process.

\begin{figure}[htbp]
\begin{minipage}[t]{0.50\textwidth}
\centering
\includegraphics[width=0.99\textwidth]{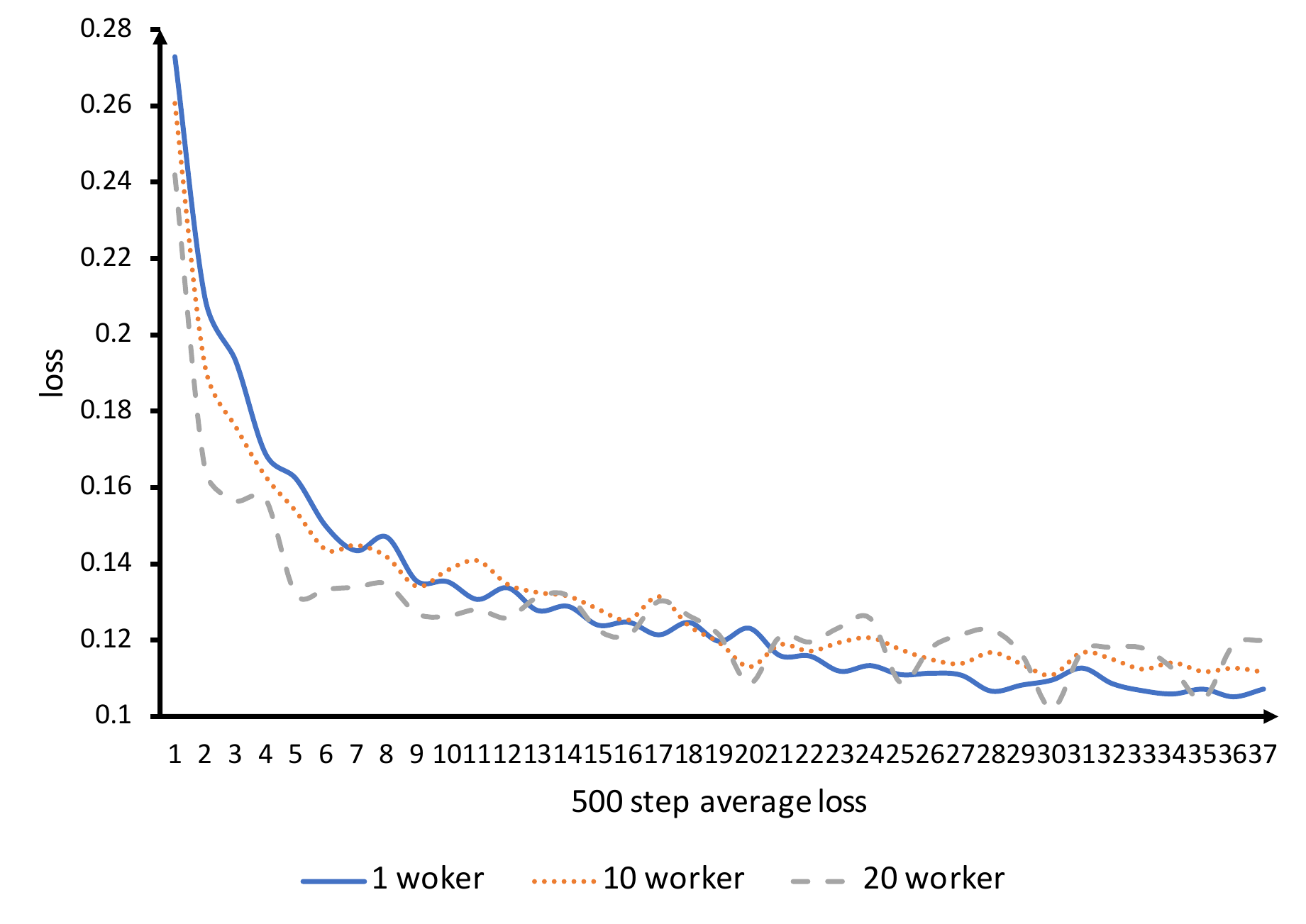}
\caption{Loss curve varying according to worker num}
\label{fig:loss_worker_num}
\end{minipage}%
\end{figure}

\begin{figure}[htbp]
\begin{minipage}[t]{0.50\textwidth}
\centering
\includegraphics[width=0.99\textwidth]{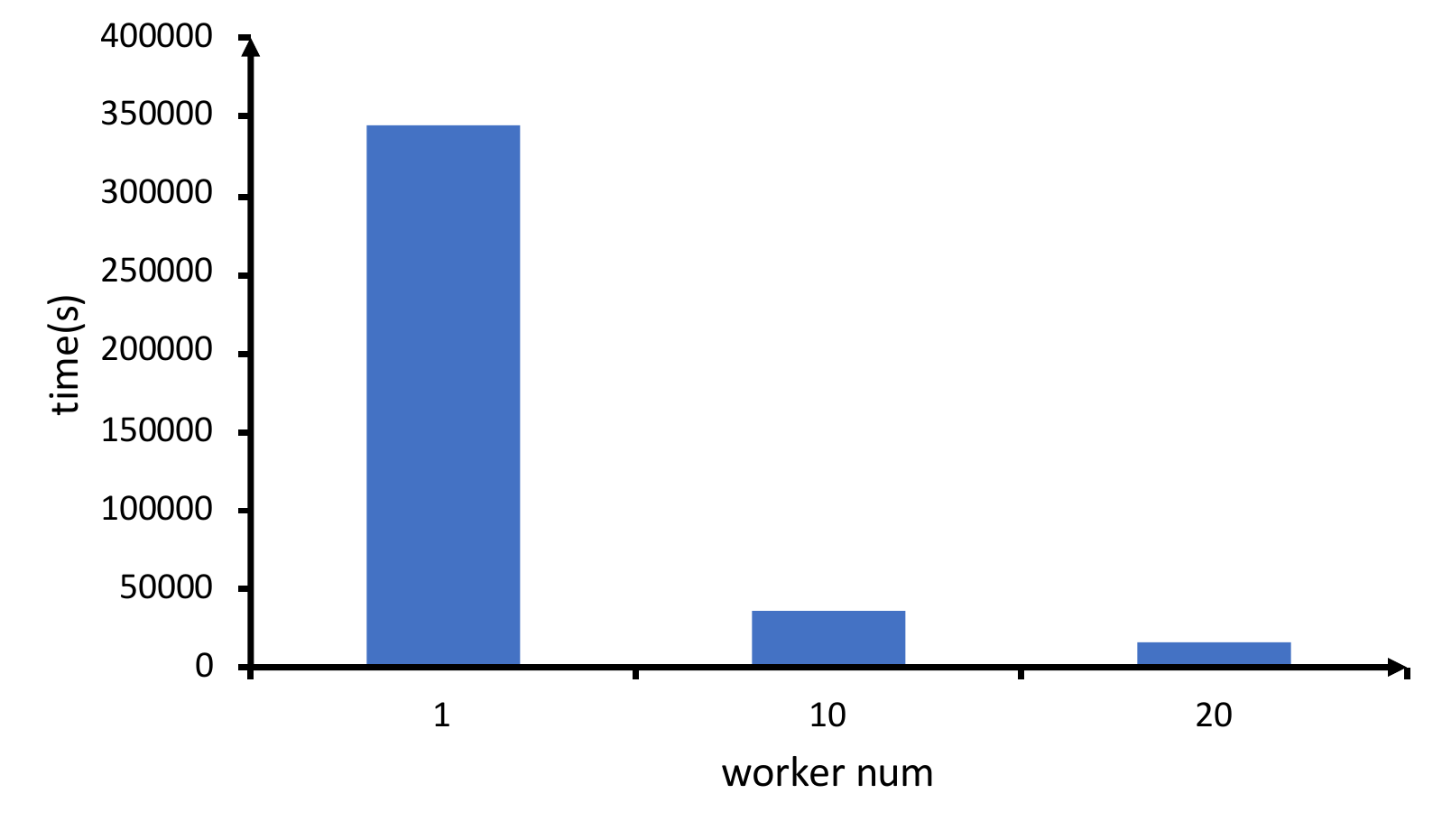}
\caption{Time according to worker num}
\label{fig:time_workerNum}
\end{minipage}%
\end{figure}

As presented in Fig.~\ref{fig:loss_worker_num} and Fig.~\ref{fig:time_workerNum}, we find that, with more worker numbers, the training process will speed up and converge to about the same level or just slightly worse than model with less workers. It's reasonable that with more workers, the performance decrease slightly as we adopt asynchronous optimization strategy during training process. We think it's acceptable when applied our model to real-world tasks with large scale data by trading off time cost and performance.

\paragraph{Offline inference efficiency} By adopting model-split strategy, the offline inference process is quite efficient compared with original inference process with Tensorflow \cite{b42}. It's takes about 15-20 hours for inferring the whole Friendship dataset with original inference process, while with model-split strategy, it only takes about 5 hours. It's worth noting that, the more dense the graph is, the better efficiency we will get with model-split strategy.

\section{Conclusion}
In this work, we propose a scalable framework for link prediction at Alipay. Based on subgraphs, our model is potential to jointly learn link prediction and node embedding on graphs with billions of nodes and edges without using any extra label information on nodes or edges.  Benefit from the shuffling and sampling strategies, our model is able to approximate distributions of positive and negative links over the whole training dataset. Moreover, the proposed model-split strategy helps accelerate the speed of inference process significantly. Experiments show that our approach works consistently well on various real-world problems. In future, we are interested in applying our framework on heterogeneous networks based on meta-path negative sampling methods.

\section{Acknowledgment}
We would like to thank our colleagues from Alipay, MY-bank and the infrastructure team of Ant Financial, who have offered great help in the devolopment and applicaiton of DSSLP. Furthermore, we like to thank Yang Shuang and Yuan Qi for their unwavering support of this project.

\vspace{12pt}
\color{red}

\end{document}